\documentstyle[PASJadd]{PASJ95}
\draft

\begin{document}

\title{Optical Number Count Estimation of IRIS
Far-Infrared Survey of Galaxies}

\author{Hiroyuki {\sc Hirashita},\thanks{%
Research Fellow of the Japan Society for the Promotion of Science}
\ Tsutomu T. {\sc Takeuchi},$^{*}$ 
and Kouji {\sc Ohta}
\\[12pt]
{\it Department of Astronomy, Faculty of Science, Kyoto University,
Sakyo-ku, Kyoto 606-8502}\\
{\it E-mail(HH): hirasita@kusastro.kyoto-u.ac.jp}\\  
and \\
Hiroshi {\sc Shibai}
\\[12pt]
{\it Department of Physics, Nagoya University, Chikusa-ku, Nagoya
464-8602}}

\abst{Infrared Imaging Surveyor (IRIS) is a satellite which will be
launched at the beginning of 2003. One of
the main purposes of the IRIS mission is an all-sky survey in the
far-infrared region with a flux limit much deeper than that of {IRAS}.
Detection of a large number of galaxies
($\sim \mbox{several}\times 10^6$ in the whole
sky) is expected in this survey.
We investigated the expected optical and near-infrared (NIR) number
counts of galaxies detected by the far-infrared
scanner (FIS) of IRIS (hereafter, IRIS galaxies) and possibility of
optical and NIR follow-up of them. The spectral energy
distribution and the luminosity
function of the IRIS galaxies are modeled based on the
properties of galaxies observed by IRAS.
The IRIS galaxies are divided into two populations
according to their infrared luminosities ($L_{\rm IR}$);
normal spirals ($L_{\rm IR}<10^{10}\LO$) and starbursts
($L_{\rm IR}>10^{10}\LO$). The expected number counts of IRIS
galaxies for both of the populations are calculated in $B$
and $H$ bands. We show that about 60 normal galaxies and about 80
starburst galaxies are detected per square degree
in both of the two bands, when galaxy evolution is not taken into
account. All of the normal population of
IRIS galaxies are located at the redshift $z\ltsim 0.1$. As for the
starburst population, we also calculated number of galaxies with a
simple model of evolution.
The total number of starburst population predicted
by the evolution model is larger by 20\% than that expected from
the non-evolution model. In the evolution model, the numbers of
low-$z$ ($z<1$), intermediate-$z$ ($1<z<3$), and high-$z$ ($z>3$)
galaxies are 100, 20, and 0.2 per square degree, respectively.
}

\kword{Galaxies: evolution --- Infrared: spectra --- Interstellar:
dust}

\maketitle
\thispagestyle{headings}

\section
{Introduction}

There have been a number of advances in understanding the galaxy
evolution over the last
few years by optical redshift surveys of galaxies (e.g.,
Lilly et al.\ 1995; Cowie et al.\ 1996; Ellis et al.\ 1996;
Hammer et al.\ 1997;
Heyl et al.\ 1997; Small et al.\ 1997).
These surveys have been revealing the star
formation history of the Universe up to $z\sim 1$.
The star formation rate around $z= 1$
is found to be several times larger than that at $z=0$. The
Lyman-break technique also provides the star formation properties
of galaxies at around $z= 3$ based on the rest ultraviolet
(UV) light (e.g., Steidel et al.\ 1996) and the star formation rate
in the Universe
seems to have
a peak at $z\sim 1$--2 (Madau et al.\ 1996). However,
optical and UV light may severely suffer from extinction by dust.

The far-infrared (FIR) is another important wavelength to trace
the star
formation history of the Universe, because the energy absorbed by
dust at UV and optical
regions is re-emitted in FIR and
extinction in FIR is very small. Furthermore, for starburst galaxies,
bulk of the energy  is released in FIR; the survey by the Infrared
Astronomical Satellite (IRAS) discovered hundreds of
galaxies emitting well over 95\% of their total luminosity in the
infrared (e.g., Soifer et al.\ 1987a). Thus, surveys only in UV and
optical
wavelengths are not sufficient
to clear up  the star formation history of the Universe: We need to
investigate the star formation history also in FIR bands.

In the FIR wavelengths, the galaxy evolution is discussed based on
the number count of the IRAS all-sky survey with a flux limit of
$\sim 1$ Jy at 60 $\mu$m. Hacking et al.\ (1987) and Saunders et al.\
(1990) showed that the observed counts of faint 60-$\mu$m sources are
about twice as high as
the non-evolutionary model prediction, suggesting the presence of
evolution of the sources. However, since the depth of the IRAS survey
is only $z\sim 0.1$ in median (Sanders, Mirabel 1996), deeper survey
is indispensable to explore the higher redshift Universe. A
survey much deeper than the IRAS survey was performed with
the Infrared Space Observatory (ISO; Kessler et al.\ 1996) at the
Lockman Hole with a flux limit of 45 mJy at 175 $\mu$m
(Kawara et al.\ 1998). According to Kawara et al.\ (1998),
the surface density of the sources brighter than 150 mJy at 175
$\mu$m agrees with the model by Guiderdoni et al.\ (1997), who took
into account the burst of star formation whose timescale of gas
consumption ($\sim 1$ Gyr) is ten times smaller than that observed
in normal disk galaxies (Kennicutt et al.\ 1994).
Although this ISO survey is much deeper
than the IRAS survey, it covers only a small solid angle of the sky
(1600 square arcmin) and much wider coverage is clearly necessary in
the next step to obtain a huge  sample.

The Infrared Imaging Surveyor (IRIS; Astro-F\footnote{Astro-F is the
project name of IRIS}) will be launched at
the beginning of the year 2003. A far-infrared scanner (FIS) as well
as a near- and
mid-infrared camera (IRC) is planned to be
on-board. A point source survey in the whole sky will be carried
out by FIS with a depth of 15--50 mJy at 50--180 $\mu$m
(Kawada et al.\ 1998).
Since the flux limit (table 1) is more than 20 times deeper
than that of IRAS, the IRIS survey is expected to detect an
enormous number of galaxies ($\sim \mbox{several}\times 10^6$ in the
whole sky; Takeuchi et al.\ 1998, hereafter T98). This
survey will contribute to studying the star formation history
and detecting primeval galaxies at high redshift (T98). The survey,
however, will not be able to determine the precise redshift
(distance) of the detected objects. In order to obtain redshift and
to know natures of the detected sources (e.g., Galactic
objects, star-forming galaxies, or AGNs), the optical or
near-infrared (NIR) follow-up observations
are indispensable. The follow-up
will enable us to obtain a systematic and homogeneous large
database more useful to study the evolution of star-forming galaxies
and AGNs as well as properties of large-scale structures.
In the optical and NIR wavelengths, the SUBARU
telescope will have been available when the IRIS survey starts.
Hence, the instrumental conditions for the follow-up
will be excellent and the follow-up of the IRIS survey will be a
timely project.


In this paper, we study the feasibility and strategy of the follow-up
of the IRIS survey using the models and calculations by T98, in
which the source count by IRIS is predicted based on the local
FIR luminosity function and the empirical spectral
energy distributions (SEDs) of nearby galaxies in the
FIR--submillimeter wavelengths, with the assumptions of simple
functional forms for
the evolution (see also Beichman, Helou 1991;
Pearson, Rowan-Robinson 1996). The models adopted
to calculate FIR number count by T98 are reviewed in section
2. Then,  we will present expected
number count of galaxies detected by IRIS (hereafter referred to
as IRIS galaxies) at
optical ($B$) and NIR ($H$) bands in section 3. Finally in section 4,
we summarize the feasibility and describe the strategy of the
follow-up of the IRIS survey.
Throughout this paper, we adopt the Hubble constant of
$H_{0}=75~\rm km~s^{-1}~Mpc^{-1}$ and the deceleration parameter of
$q_{0}=0.1$ (the density parameter $\Omega_0=0.2$ and the cosmological
constant $\Lambda =0$) unless otherwise stated.


\section{Model of FIR Number Count}

In this section, we review the model used by T98 to calculate
the expected number count of the IRIS survey in FIR bands.

\subsection{Spectral Energy Distribution}

The assumed SEDs in T98 consist of two components, cool cirrus
and hot starbursts, following Rowan-Robinson, Crawford (1989).
The cirrus component represents the dust heated mainly by
late-type stars, while the starburst component stands for
emission from dust heated by hot early-type stars associated
with intense starbursts (e.g., Sanders, Mirabel 1996). 
The model spectrum of cirrus component is taken from the Galactic
interstellar dust model by D\'{e}sert et al.\ (1990). In T98, it was
assumed that galaxies with $L_{\rm IR}<10^{10}~\LO$ have cirrus
component only, where $L_{\rm IR}$ is defined as
infrared luminosity integrated in the wavelength range from 3 $\mu$m
to 1 mm.
Galaxies with larger $L_{\rm IR}$ have both the cirrus and starburst
components; the infrared luminosity in excess of $10^{10}~\LO$ is
assumed to
come from the starburst component. The spectrum of starburst
component $L_{{\rm s}\nu}$ is composed of two-temperature modified
blackbody radiation:
$L_{{\rm s}\nu}=\alpha\nu B_\nu (T_{\rm cool})+\beta\nu B_\nu
(T_{\rm hot}),$
where $B_\nu (T)$ is the Planck function with a temperature $T$ and
both $\alpha$ and $\beta$ are normalizing constants
(Rowan-Robinson, Crawford 1989). Here, the temperature of the two
components are given by
$T_{\rm cool}=60({L_{\rm s}}/{10^{11}~\LO})^{0.1}\,{\rm K}$
and
$T_{\rm hot}=175({L_{\rm s}}/{10^{11}~\LO})^{0.1}\,{\rm K},$
where $L_{\rm s}$ is the luminosity of starburst component
integrated in the range from 3 $\mu$m to 1 mm
(Beichman, Helou 1991; see also Rieke, Lebofsky 1986; Helou 1986;
Soifer et al.\ 1987a). Normalizing constants $\alpha$ and $\beta$
are determined so that 70\% of the starburst component comes from
the cool component and 30\% from the hot component
(Beichman, Helou 1991). The model SED is scaled
properly to yield the given $L_{\rm IR}$.
In the model, AGNs are not considered, since
the number of AGNs is expected to be negligible compared with
the number  of normal and starburst galaxies (Beichman, Helou
1991). Even if the evolution of AGNs is considered, this will also
be the case (Pearson 1996).

\subsection{Local Luminosity Function}

The local FIR luminosity function of galaxies is derived from
the IRAS data (Soifer et al.\ 1987b). Since Soifer et al.\ (1987b)
used the FIR luminosity at 60 $\mu$m, T98 converted
it into their $L_{\rm IR}$ by using their model SEDs. The analytical
fitting to the data leads to the following double-power-law form
of the luminosity function:
\begin{eqnarray}
\log [\phi_0(L_{\rm IR})]
=\left\{
\begin{array}{ll}
7.9-1.0\log\left({L_{\rm IR}}/{\LO}\right) & \mbox{for} \
10^8\LO <L_{\rm IR}<10^{10.3}\LO ; \\
17.1-1.9\log\left({L_{\rm IR}}/{\LO}\right) & \mbox{for} \
10^{10.3}\LO <L_{\rm IR}<10^{14}\LO ; \\
\mbox{no galaxies} & \mbox{otherwise},\label{lf}
\end{array}
\right.
\end{eqnarray}
where $\phi_0$ is the number density of galaxies in
Mpc$^{-3}$ dex$^{-1}$ at $z=0$.

\subsection{Survey Limit}

IRIS is a survey-type infrared telescope in space. Its major goal is
to achieve the whole sky point source survey in the far-infrared
wavelength region with an order of magnitude higher sensitivity and
higher spatial
resolution as well as with a longer wavelength photometric band
than those of the IRAS survey.

The flux limit of the all-sky IRIS survey is estimated by Kawada
et al.\ (1998) as 20 mJy at
$\sim 50$ $\mu$m, 15 mJy at $\sim 70\,\mu$m, 30 mJy at
$\sim 120\,\mu$m, and 50 mJy at
$\sim 150\,\mu$m (table 1). In the shorter wavelength region
of 50--110 $\mu$m, the sensitivity is limited by both internal and
background noises,
while in the longer wavelength range of 110--170 $\mu$m, the detection 
limit is constrained by the confusion
effects of the interstellar cirrus in the Galaxy.
Much better point source detection limits can be
expected in limited sky areas near the ecliptic poles, where the
survey scan will be repeated more than a hundred times.
In this paper, we adopt a conventional
value of the point source detection limit expected for the all-sky
survey.

\subsection{Redshift Distribution of IRIS Galaxies}

Assuming the above SED and luminosity function of galaxies, T98
calculated the redshift distribution of IRIS galaxies which were
regarded as point sources there. The redshift distribution per
unit solid angle and unit redshift is formulated as
\begin{eqnarray}
\frac{{\rm d}^2N}{{\rm d}\Omega{\rm d}z}=
\frac{{\rm d}^2V}{{\rm d}\Omega{\rm d}z}
\int_{L_{\rm lim}(z)}^\infty \phi (L_{\rm IR}',\, z)\, {\rm d}\log
L_{\rm IR}'
\label{zdist},
\end{eqnarray}
where the effect of evolution is included in the luminosity function
at $z$, $\phi (L_{\rm IR}',\, z)$, and the comoving volume
element per str and per $z$ is denoted as
${{\rm d}^2V}/{{\rm d}\Omega{\rm d}z}$ ($\Omega$ is the solid
angle). In equation (\ref{zdist}),
$L_{\rm lim}(z)$ is the limiting luminosity for the source at
$z$. The limiting luminosities are $10^{11}\LO$ at $z\sim 0.2$,
$10^{12}\LO$ at $z\sim 1$, and $10^{13}\LO$ at $z\sim 3$.
$L_{\rm lim}(z)$ becomes $10^{14}\LO$ at $z\sim 5$ so that no
galaxies are detected in $z>5$ due to upper luminosity cutoff of
the luminosity function (equation \ref{lf}). The comoving volume
element can be
expressed in terms of cosmological parameters as
\begin{eqnarray}
\frac{{\rm d}^2V}{{\rm d}\Omega{\rm d}z}=
\frac{c}{H_0}\frac{d_{\rm L}^2}{(1+z)^3\sqrt{1+2q_0z}},
\label{kolb}
\end{eqnarray}
(e.g., Kolb \& Turner 1994), where $c$ is the speed of light and
$d_{\rm L}$ is the luminosity distance:
\begin{eqnarray}
d_{\rm L}=\frac{c}{H_0q_0^2}[zq_0+(q_0-1)(\sqrt{2q_0z+1}-1)]\; .
\end{eqnarray}

The resultant number count per $z$ in the whole sky is shown in
figure 1 for 50-, 70-, 120-, and 150-$\mu$m surveys in the case of no
evolution. 
The total number count of the IRIS galaxies integrated over all the
redshift range is several times $ 10^6$ in the whole sky at each
wavelength. The 120-$\mu$m survey is the deepest ($z\sim 5$).
Both of the 70- and 150-$\mu$m surveys reach $z\sim 4.5$ and the
50-$\mu$m survey reaches $z\sim 3.5$. Hence, the 120-$\mu$m survey is
preferable to detect high-$z$ galaxies. As for lower redshift region,
 the number at 150 $\mu$m
is one order of magnitude smaller than the others in the range of
$0.2\ltsim z\ltsim 1$, and the number
count at 50 $\mu$m is one order of magnitude smaller than the other
bands at $z\ltsim 0.2$. The redshift distribution of the IRIS
galaxies is discussed further in T98.

\subsection{Treatment of Galaxy Evolution}

For galaxy evolution models, T98 treated pure luminosity
evolution and
pure density evolution. In this subsection, we briefly review
the method to include the evolution effect.

\subsubsection{Pure luminosity evolution}

The pure luminosity evolution in T98 means that the luminosities
of galaxies change as a function of redshift with the functional
form of the luminosity function fixed (equation \ref{lf}).
The effect of the luminosity evolution of galaxies is so modeled
that the luminosities of galaxies increase by a factor $f(z)$ at
a redshift $z$: 
$L_{\rm IR}(z)=L_{\rm IR}(z=0)f(z)$.
The function $f(z)$ is empirically given in T98 as follows:
\begin{eqnarray}
f(z)=\exp\left[Q\frac{\tau(z)}{t_{\rm H}}\right] ,\label{fz}
\end{eqnarray}
where $Q$, $\tau (z)$ and $t_{\rm H}$ are, respectively, a parameter
defining the magnitude of evolution, look-back time as a function
of $z$, and the Hubble time $1/H_0$. 

\subsubsection{Pure density evolution}

The pure density evolution in T98 means that only the comoving
number density of galaxies changes as a function of redshift with
the luminosities of the galaxies fixed.
The density evolution is so modeled by using a function $g(z)$
that the comoving number density at $z$ is $g(z)$ times
larger than that at $z=0$: $\phi (z,\, L_{\rm IR})=\phi_0
(L_{\rm IR})g(z)$, where $\phi (z,\, L_{\rm IR})$ is the
luminosity function in the comoving volume at the redshift of $z$.
For $g(z)$, T98 adopted
the functional form similar to $f(z)$ defined in equation (\ref{fz}):
\begin{eqnarray}
g(z)=\exp\left[ P\frac{\tau (z)}{t_{\rm H}}\right] ,
\end{eqnarray}
where $P$ is a parameter representing the magnitude of the evolution.

\subsubsection{Determination of parameters and observational
constraints}

In T98, each of
the evolutionary parameters ($P$ and $Q$) is determined by comparison
with the IRAS extragalactic source count data. Excluding four
statistically poorest points, the values fit best to the data of 
IRAS Point Source Catalog (Joint IRAS Science Working Group 1985)
and of Hacking, Houck (1987) are $P=2.7$ or
$Q=1.4$ when $q_0=0.1$.
Recent ISO result at 175 $\mu$m by Kawara et al.\ (1998) is consistent
with the model prediction of T98. The cosmic infrared background
radiation predicted by the model of T98 is consistent with DIRBE and
FIRAS results (Puget et al.\ 1996; Fixsen et al.\ 1998; Hauser et al.\
1998) in the FIR region. We note that the submillimeter background
radiation predicted by T98 is a few--ten times larger than the 
DIRBE and FIRAS results. The redshifted
dust emission whose peak is located at 30--100 $\mu$m at the rest 
frame of the galaxy (figure 1 of T98) largely contributes to
the submillimeter background.
Thus, constraint on the galaxy evolution in the high-redshift
Universe from the submillimeter
background is suggested to be useful for future works.
The model is also compared with the SCUBA
data (Smail et al.\ 1997; Hughes et al.\ 1998; Barger et al.\ 1998)
in appendix of T98.

\section{Number Count of IRIS Galaxies in Optical and NIR
Wavelengths}

In this section, we estimate
the number of the IRIS galaxies for $B$ ($\lambda =4400$ \AA)
and $H$ ($\lambda =16500$ \AA) bands.
Galaxy evolution
is not taken into account at first. The effects of the evolution are
considered later in this section.

\subsection{Method for Calculation}

We calculate the expected numbers of the IRIS galaxies at $B$ and
$H$ bands in the following way:

[1] Two populations for the IRIS galaxies are assumed;
starburst population and normal spiral population.
We define the starburst population as galaxies whose
$L_{\rm IR}$ exceeds $10^{10}~\LO$. This classification of
the populations corresponds to two different compositions of
FIR SEDs in section 2.1.

[2] For the UV-to-FIR SED of each population, we used averaged
SEDs of sample galaxies in Schmitt et al.\ (1997). They collected
the nearby galaxies from the catalog of ultraviolet IUE spectra
(Kinney et al.\ 1993, 1996), whose
ground-based spectra observed with apartures matching that of IUE
($10''\times 20''$) were available. The sample contains 6
normal spirals and 26 starburst galaxies. The starburst
galaxies are
divided into two categories; the high-reddening sample
(15 galaxies) and the low-reddening
sample (11 galaxies) at $E(B-V)=0.4$. The reddening was calculated
in Calzetti et al.\ (1994) from Balmer decrement with Seaton's
reddening law (Seaton 1979).
In this paper, we used the high-reddening sample, so
that the UV or optical luminosity density converted from the FIR
luminosity density gives fainter estimation. The conversion
of $L_{\rm IR}$ to the luminosity for each band (at frequency $\nu$)
is made for the two populations using the averaged SEDs as
${\cal L}_\nu /{\cal L}_{60\,\mu{\rm m}}=\alpha_\nu$, where
${\cal L}_\nu =\nu L_\nu$. We should note that
${\cal L}_{60\,\mu{\rm m}}$ and $L_{\rm IR}$ are related to each
other (section 2.2).
The value of $\alpha_\nu$, which is assumed to be a function of
$\nu$ only, is listed in table 2 for each population and wavelength.
The ratio of luminosity density at $\nu$ to that
at $60\,\mu$m is kept constant for each population. 
We should keep in mind that the scatter of $\alpha_\nu$ is
large (an order of magnitude) in observational data.
 Since our estimation is made for FIR-selected sample, use of
SEDs in Spinoglio et al.\ (1995), in which the sample was selected
at 12 $\mu$m, is better. However, we need UV data to evaluate
the optical magnitude of galaxies at high redshift, and thus we
used the data in Schmitt et al.\ (1997).
It is worth noting that
the FIR-to-optical SEDs of 12-$\mu$m sample in Spinoglio et al.\
(1995; their figure 11) is consistent with the value of $\alpha_\nu$
in table 1 within an order of magnitude, typical scatter of the data
in Schimitt et al.\ (1997).

[3] The luminosity function $\phi_0({\rm AB}_\nu)$
(Mpc$^{-3}$ mag$^{-1}$) is made based on Soifer et al.\ (1987b) by
using the above conversion of luminosity density. The definition
of the AB magnitude (${\rm AB}_\nu$) is given by Oke \& Gunn (1983);
${\rm AB}_\nu\mbox{(mag)} = -2.5\log f_\nu
(\mbox{erg cm}^{-2}~{\rm s}^{-1}~\mbox{Hz}^{-1})- 48.594\,
(B={\rm AB}+0.2;\, H={\rm AB}-1.4)$.
The $B$-band luminosity function is described as follows. For the
normal population,
\begin{eqnarray}
\log\phi_0~(\mbox{Mpc}^{-3}~\mbox{mag}^{-1}) = 5.08+0.38
{\rm AB}_{B,{\rm abs}}, \ \ \
 (-19.8<{\rm AB}_{B,{\rm abs}}<-14.8).
\end{eqnarray}
where $\phi_0$ is the local number density of the IRIS galaxies per
magnitude, and the subscript ``abs'' refers to absolute magnitude. The
upper and lower luminosity correspond to $L_{\rm IR}=10^{10}\LO$
(${\cal L}_{60\,\mu{\rm m}}=10^{9.2}\LO$)
and $L_{\rm IR}=10^8\LO$ (${\cal L}_{60\,\mu{\rm m}}=10^{7.2}\LO$),
respectively. For the starburst population, double-power-law fitting
is executed as follows:
\begin{eqnarray}
\lefteqn{\log\phi_0~(\mbox{Mpc}^{-3}~\mbox{mag}^{-1})} \nonumber \\
& = & \left\{
\begin{array}{ll}
0.02+0.16
{\rm AB}_{B,{\rm abs}} & (-18.0<{\rm AB}_{B,{\rm abs}}<-15.8), \\
12.07+0.83{\rm AB}_{B,{\rm abs}} &
(-25.6<{\rm AB}_{B,{\rm abs}}<-18.0).
\end{array}
\right.
\end{eqnarray}
Here, the boundary values for the luminosity
correspond to $L_{\rm IR}=10^{14}\LO$
(${\cal L}_{60\,\mu{\rm m}}=10^{13.2}\LO$;
${\rm AB}_{B,{\rm abs}}=-25.6$),
 $L_{\rm IR}=10^{10.3}\LO$ (${\cal L}_{60\,\mu{\rm m}}=10^{10.1}\LO$;
${\rm AB}_{B,{\rm abs}}=-18.0$)
and $L_{\rm IR}=10^{10}\LO$ (${\cal L}_{60\,\mu{\rm m}}=10^{9.2}\LO$;
${\rm AB}_{B,{\rm abs}}=-15.8$).
For $H$ band, the luminosity function of the normal population is
\begin{eqnarray}
\log\phi_0\, (\mbox{Mpc}^{-3}~\mbox{mag}^{-1})=6.00+0.38
{\rm AB}_{H,{\rm abs}} \ \ \
 (-22.2<{\rm AB}_{H,{\rm abs}}<-17.2),
\end{eqnarray}
while the starburst luminosity function is
\begin{eqnarray}
\lefteqn{\log\phi_0~(\mbox{Mpc}^{-3}~\mbox{mag}^{-1})} \nonumber \\
& = & \left\{
\begin{array}{ll}
0.10+0.15{\rm AB}_{H,{\rm abs}} &
(-19.6<{\rm AB}_{H,{\rm abs}}<-17.4), \\
13.40+0.83{\rm AB}_{H,{\rm abs}} &
(-27.2<{\rm AB}_{H,{\rm abs}}<-19.6).
\end{array}
\right.
\end{eqnarray}
The fraction of starburst
galaxies is 2\% of total number of field galaxies at
${\rm AB}_B\sim -20$ mag (see e.g., Small et al.\ 1997 for the
number density of field galaxies).


[4] The $K$-corrections, $K(z)$ is expressed by 5-order polynomials
fitted to the data of SEDs in Schmitt et al.\ (1997) as
$K(z)=a_1z+a_2z^2+a_3z^3+a_4z^4+a_5z^5.$ The results of the fitting
are presented in table 3.
The residual of the fitting of $K(z)$ is less than $0.2$ mag in
most of the considered $z$ range and 0.5 mag in the worst case; the
values are less than the scatter of the
luminosity density of the sample in Schmitt et al.\ (1997).


[5] Finally, the number count of IRIS galaxies per square degree is
calculated according to the following formula:
\begin{eqnarray}
N(<{\rm AB}_\nu) & = & \oint
{\rm d}\Omega\int_0^{\infty}
{\rm d}z\int_{{\rm AB}_{\nu,{\rm lim}}(z)}^\infty
{\rm d}\,{\rm AB}_\nu\, \phi ({\rm AB}_\nu )
\frac{{\rm d}^2V}{{\rm d}z\,{\rm d}\Omega}\; , \label{gnc}
\end{eqnarray}
where ${\rm AB}_{\nu,{\rm lim}}(z)$ is the limiting magnitude
corresponding to $L_{\rm lim}(z)$ (section 2.3).

\subsection{Results}

The cumulative number counts of IRIS galaxies detected at 120 $\mu$m
are presented in figures 2a ($B$ band) and 2b ($H$ band). These
figures show that number of IRIS galaxies brighter than
AB$_B\sim 19$ mag (or AB$_H\sim 16$ mag) increases as
magnitude increases with a slope of 0.6 (dex mag$^{-1}$),
which is the value for the no evolution case in the Eucledian
geometry. The figures also show that no
IRIS galaxies are detected in the magnitude region fainter than
${\rm AB}_B\sim 22$ mag (or ${\rm AB}_H\sim 21$ mag) for starbursts
and ${\rm AB}_B\sim 19$ mag (or ${\rm AB}_H\sim 16$ mag) for
normal spirals. Each of these
magnitude corresponds to optical flux density of each population
which is detected at the flux limit of the IRIS survey. About 60
normal galaxies and 80 starbursts per square degree are detected
within this limit.
The redshifts of normal spirals are less than $0.1$ (T98) as expected
from the slope of the counts. In
figures 3a and 3b, we also present the number count of starbursts
in various redshift
range ($z<0.2,\, 0.2<z<1,\, 1<z<3,$ and $3<z$) in the case of
the 120-$\mu$m survey. About 90\%
of the starbursts detected by IRIS are located at the redshift of
$z<0.2$ (40\% at $z<0.1$ and 50\%
at $0.1<z<0.2$). 
For $z\ltsim 3$, the IRIS galaxies are
brighter than ${\rm AB}_B\sim 22$ (or ${\rm AB}_H\sim 20$),
corresponding to the detection limit of IRIS. One high-$z$ ($z>3$)
galaxy exists per $\sim 10$ square degrees. 


We used the high-reddening SED of Schmitt et al.\ (1997) as
that of starburst galaxies (section 3.1). The UV-to-FIR ratio of
luminosity density is about 2 times larger for the low-reddening
SED than
for the high-reddening SED. Thus, the optical counterpart of
IRIS galaxies may be brighter than expected in this paper
by two times ($\sim 0.8$ mag), if the dust extinction is lower.
Even if the deceleration parameter of the Universe is not 0.1 but
0.5, there is no significant change in the detected number at
$z\ltsim 2$. The number increases by 20 \% around $z=3$.

\subsection{Effects of Galaxy Evolution}

Guiderdoni et al.\ (1997) calculated number count of galaxies in
FIR wavelength
based on the models constructed through detailed physical processes
related to the evolution of galaxies (see also Franceschini et al.\
1994). However, there is a great deal of theoretical uncertainty on
the physical processes governing galaxy formation and evolution.
Thus, we adopt ``empirical approach'' (Ellis 1997) based on
the luminosity function and SEDs observed in the local universe.

We, here, investigate the pure luminosity evolution in T98. Their
model is reviewed in subsection 2.5.
The evolution with the parameter $Q=1.4$, which T98
determined by using IRAS extragalactic source count data,
is examined in this
paper. This model shows the luminosity evolution of a factor
of 1.9 at $z=1$, 2.4 at $z=2$, and maintain this evolutionary factor
in $z\gtsim 2$.
We assume for simplicity that the luminosity of galaxies
evolves in such a way that the ratio of FIR to optical luminosity is
kept constant.

Figures 4a and 4b present the effect of the evolution for the
starburst population. Since almost all of the IRIS galaxies reside in
low-$z$ area, the effect of galaxy evolution is not
significant (increase of about 20\%). We also present the number count
for various range of $z$ in figures
5a and 5b. Comparing figures 3 and 5, we see that the effect of
evolution is significant for high-$z$ galaxies;
the number of
low-$z$ ($z<1$), intermediate-$z$ ($1<z<3$), and high-$z$ ($z>3$)
galaxies are 100, 20, and 0.2 per square degree, respectively.



The intensity of cosmic infrared background radiation (e.g.,
Hauser 1995) also constrains the magnitude of galaxy evolution. Since
$Q=1.4$ is almost upper limit for the magnitude, stronger
evolutions with our model break the constraint (T98).
The density evolution of luminosity function in T98 gives almost
the same results, since the parameter for the density evolution
is determined from the same calibration as the luminosity evolution
(see T98 for details).

\section{Summary and Discussion}


We adopted the ``empirical approach'' model
to predict the optical and NIR number count of galaxies expected
to be detected by IRIS. The model is based on the local observed
data of IRAS and is the extension of the IRAS results to high-$z$
universe. According to our model, such IRIS galaxies
have magnitude ${\rm AB}_\nu\ltsim 21$ at $B$ and $H$ bands.
The expected number of IRIS galaxies at $B$ and $H$ bands
per square degree is estimated as 80 for starburst galaxies
and 60 for normal spirals. The value is about 10\% of optical/NIR
number count at the same magnitude.
As for the redshift distribution, almost all of the normal galaxies
are located at $z\ltsim 0.1$, and 40\% of the starburst galaxies are
at $z< 0.1$, 50\% at $0.1<z<0.2$, 10\% at $0.2<z<1$, and 1\% at
$z>1$.

By considering the results obtained above together with the
effects of the evolutions, scientific targets
of optical/NIR follow-up observations of IRIS galaxies would
be twofold. One is to trace star formation properties as well
as a large scale structure of the Universe up to $z\sim 1$.
Environmental effects on star formation in galaxies will be an
important issue. Another target is to find extreme starburst
galaxies in the high-$z$ Universe, which are in an early stage of
galaxy evolution.

IRIS is expected to determine the position of a FIR source with an
accuracy of $5''$ (Kawada et al.\ 1998), which is estimated based on
the accuracy of telescope pointing and of fitting to a beam profile.
The observed number counts of
galaxies in $b_{\rm J}$ and $K$ bands show that about $10^{-2}$
galaxies exist in a $5''\times 5''$ field at an AB magnitude of 21
mag (e.g., Broadhurst et al.\ 1992). This means that the chance
coincidence between IRIS
galaxies and normal optical galaxies is negligible within this
magnitude limit, and thus
we can select optical counterpart of IRIS galaxies almost uniquely.
Since the expected redshifts of most of such IRIS galaxies are low
($z\ltsim 1$), optical spectroscopy will be good enough to
know redshifts and natures of the sources. Considering that the
number density of the IRIS galaxies brighter than ${\rm AB}\simeq 21$
mag is $\sim 100$ deg$^{-2}$, a
multi object spectrograph with a wide field of view (such as fiber
multi object spectrograph) equipped to a 4--8-m class telescope is
the best instrument to follow
up the IRIS survey. The obtained database will be used to trace
star formation history and large scale structures up to $z\sim 1$.

It will be very inefficient to find high-$z$ ultra-luminous
FIR galaxies in such a survey described in the previous
paragraph because of its very low surface density. We need
another approach to obtain a high-$z$ ultra-luminous FIR
sample. It is expected that we can make a rough estimate of the
redshifts of IRIS galaxies based on a FIR color-color diagram of
the sources by
using three of IRIS bands as discussed in T98. Their basic idea is
based on the fact that the
peaks in spectra of thermal radiation from the heated dust at
high $z$ is redshifted to a larger wavelength. After the selection
of the high-$z$ galaxies by their FIR colors, we need deep
spectroscopic observations targeting these objects. Since such
high-$z$ objects
will have faint magnitude (AB\ltsim 23), optical counterparts may
not be uniquely identified; chance probability is not negligible
in this magnitude range. 
Furthermore, ``dusty''
galaxies may have fainter optical magnitude. Thus, deep multislit
optical/NIR spectroscopy or integral field unit of a field of view
of $\sim 10''$ will be required. Since making slitlets in a
5$''$--10$''$ region would be difficult, integral field unit would
be the most
efficient way to identify the FIR source. Hence, integral field
units as well as multi object spectrographs on a 4--8-m class
telescope will be powerful tools to conduct optical/NIR
follow-ups of the IRIS survey.

\par
\vspace{1pc}\par
We are grateful to the anonymous referee for useful comments
which improved our paper.
We would like to thank Profs.\ M.\ Sait\={o} and S.\ Mineshige for
continuous encouragement. We acknowledge Drs.\ H.\ Matsuhara,
M.\ Kawada and other
IRIS staff members for useful discussions and comments.
We also thank Drs.\ T.\ Yamada, H.\ Kamaya, T.\ T.\ Ishii
and T.\ G.\ Hattori for their kind helps and fruitful discussions.
Two of the authors (HH and TTT) acknowledge the Reserch Fellowship of
the Japan
Society for the Promotion of Science for Young
Scientists.


\section*{References}
\small

\re
Barger A.J.,  Cowie L.L., Sanders D.B., Fulton E., Taniguchi Y.,
Sato Y., Kawara K., Okuda H.\ 1998, Nature 394, 248
\re
Beichman C.A., Helou, G.\ 1991, ApJ 370, L1
\re
Broadhurst T.J., Ellis R.S., Glazebrook K.\ 1992, Nature 355, 55
\re
Calzetti D., Kinney A.L., Storchi-Bergmann T.\ 1994, ApJ 429, 582 
\re
Cowie L.L., Songaila A., Hu E.M., Cohen J.G.\ 1996, AJ, 112, 839
\re
D\'{e}sert F.-X., Boulanger F., Puget J.L.\ 1990, A\&A 237, 215
\re
Ellis R.S.\ 1997, ARA\&A 35, 389
\re
Ellis R.S., Colless M., Broadhurst T., Heyl J., Glazebrook K.\
1996, MNRAS 280, 235
\re
Fixsen D.J., Dwek E., Mather J.C., Bennett C.L., Shafer R.A.\
1998, ApJ accepted (astro-ph/9803021)
\re
Franceschini A., Mazzei P., De Zotti G., Danese L.\ 1994,
ApJ 427, 140
\re
Guiderdoni B., Hivon E., Bouchet F.R., Maffei B.\ 1997, MNRAS 295,
877
\re
Hacking P., Condon J.J., Houck J.R.\ 1987, ApJ 316, L15
\re
Hacking P, Houck J.R.\ 1987, ApJS 63, 311
\re
Hammer F., et al.\ 1997, ApJ 481, 49
\re
Hauser M.G.\ 1995, in Unveiling the Cosmic Infrared Background,
ed. E. Dwek, (AIP) p11
\re
Hauser M.G., et al.\ 1998, ApJ, accepted (astro-ph/9806167)
\re
Heyl J., Colless M., Ellis R.S., Broadhurst T.\ 1997, MNRAS 285,
613
\re
Helou G.\ 1986, ApJ 311, L33
\re
Hughes D.H., et al.\ 1998, Nature 394, 241
\re
Joint IRAS Science Working Group 1985, IRAS Point Source Catalog
(GPO, Washington DC)
\re
Kawada M.\ et al.\ 1998, in Infrared Astronomical Instrumentation,
Proc.\ SPIE, in press
\re
Kawara K., et al.\ 1998, A\&A 336, L9
\re
Kennicutt R.C. Jr., Tamblyn P., Congdon C.W.\ 1994, ApJ 435, 22
\re
Kessler M., et al.\ 1996, A\&A 315, L27
\re
Kinney A.L., Bohlin R.C., Calzetti D., Panagia N., Wyse R.F.G.\
1993, ApJS 86, 5
\re
Kinney A.L., Calzetti D., Bohlin R.C., McQuade K., Storchi-Bergmann,
T., Schmitt H.R.\ 1996, ApJ 467, 38
\re
Kolb E.W., Turner M.S.\ 1994, {The Early Universe}
(Addison Wesley)
\re
Lilly S.J., Tresse L., Hammer F., Crampton D., Le F\`{e}vre O.\ 1995,
ApJ 455, 108
\re
Madau P., Ferguson H.C., Dickinson M.E., Giavalisco M., Steidel C.C.,
Fruchter A.\ 1996, MNRAS 283, 1388
\re
Oke J.B., Gunn J.E.\ 1983, ApJ, 266, 713
\re
Pearson C.\ 1996, PhD Thesis
\re
Pearson C., Rowan-Robinson M.\ 1996, MNRAS 283, 174
\re
Puget J.-L., Abergel A., Bernard J.-P., Boulanger F., Burton W.B.,
D\'{e}sert F.-X., Hartmann D.\ 1996, A\&A 308, L5 
\re
Rieke G.H., Lebofsky M.J.\ 1986, ApJ 304, 326
\re
Rowan-Robinson M., Crawford J.\ 1989, MNRAS, 238, 523
\re
Sanders D.B., Mirabel I.F.\ 1996, ARA\&A 34, 749
\re
Saunders W., Rowan-Robinson M., Lawrence A., Efstathiou G.,
Kaiser N., Ellis R.S., Frenk C.S.\ 1990, MNRAS 242, 318
\re
Schmitt H.R., Kinney A.L., Calzetti D., Storchi-Bergmann T.\ 1997,
    AJ 114, 592
\re
Seaton M.J.\ 1979, MNRAS 187, L73
\re
Smail I., Ivison R.J., Blain A.W.\ 1997, ApJ 490, L5
\re
Small T.A., Sargent W.L.W., Hamilton D.\ 1997, ApJ 487, 512
\re
Steidel C.C., Giavalisco M., Pettini M., Dickinson M., Adelberger
K.L.\ 1996, ApJ 462, L17
\re
Soifer B.T., Houck J.R., Neugebauer G.\ 1987a, ARA\&A 25, 187
\re
Soifer B.T., Sanders D.B., Madore B.F., Neugebauer G., Danielson G.E.,
Elias J.H., Lonsdale C.J., Rice W.L.\ 1987b, ApJ 320, 238
\re
Spinoglio L., Malkan M.A., Rush B., Carrasco L., Recillas-Cruz E.\
1995, ApJ 453, 616
\re
Takeuchi T.T., Hirashita H., Ohta K., Hattori T.G., Ishii T.T.,
Shibai H.\ 1998, PASP submitted (T98)

\label{last}

\clearpage

\begin{table}[t]
\small
\begin{center}
Table~1.\hspace{4pt}Flux limits of the IRIS far-infrared scanner.\\
\end{center}
\vspace{6pt}
\begin{tabular*}{\columnwidth}{@{\hspace{\tabcolsep}
\extracolsep{\fill}}p{9pc}cc}
\hline\hline\\[-6pt]
Wavelength [$\mu$m] & 5$\sigma$-detection limit [mJy] \\
[4pt]\hline \\[-6pt]
50 \dotfill & 20 \\
70 \dotfill & 15 \\
120 \dotfill &  30 \\
150 \dotfill    & 50 \\ [4pt]
\hline
\end{tabular*}
\end{table}

\clearpage

\begin{table}[t]
\small
\begin{center}
Table~2.\hspace{4pt}Parameter, $\alpha$, for the conversion of
luminosities. \\
\end{center}
\vspace{6pt}
\begin{tabular*}{\columnwidth}{@{\hspace{\tabcolsep}
\extracolsep{\fill}}p{9pc}cc}
\hline\hline\\[-6pt]
Population (band) & $\alpha^*$ \\
[4pt]\hline \\[-6pt]
Starburst ($B$) \dotfill & 0.10 \\
Normal ($B$) \dotfill    & 4.0 \\
Starburst ($H$) \dotfill & 0.12 \\
Normal ($H$) \dotfill    & 10 \\ [4pt]
\hline
\end{tabular*}
\vspace{6pt}\par\noindent
$*$ $\alpha ={\cal L}_\nu /{\cal L}_{60\mu{\rm m}}$, where
${\cal L}_\nu =\nu L_\nu$ ($L_\nu$ is luminosity density of
a galaxy).
\end{table}

\clearpage

\begin{table}[t]
\small
\begin{center}
Table~3.\hspace{4pt}The coefficient $a_n$ for the $K$-correction. \\
\end{center}
\vspace{6pt}
\begin{tabular*}{\columnwidth}{@{\hspace{\tabcolsep}
\extracolsep{\fill}}p{9pc}ccccc}
\hline\hline\\[-6pt]
Population (band) & $a_1$ & $a_2$ & $a_3$ & $a_4$ & $a_5$ \\
[4pt]\hline \\[-6pt]
Starburst ($B$) \dotfill & 2.1 & $-1.2$ & 0.25 & $-2.3\times 10^{-2}$
& $7.4\times 10^{-4}$ \\
Normal ($B$) \dotfill    & 5.7 & $-2.6$ & $-0.39$ &
$-1.3\times 10^{-2}$ & $-4.6\times 10^{-4}$ \\
Starburst ($H$) \dotfill & $-0.49$ & 0.38 & $-7.0\times 10^{-2}$ &
$5.2\times 10^{-3}$ & $-1.4\times 10^{-4}$ \\
Normal ($H$) \dotfill    & $-0.22$ & 0.35 & $-3.3\times 10^{-2}$ &
$1.1\times 10^{-4}$ & $3.3\times 10^{-5}$ \\ [4pt]
\hline
\end{tabular*}
\end{table}

\clearpage
\bigskip
\begin{fv}{1}
{7cm}
{Number of the IRIS galaxies per unit $z$.
The solid, dashed, dotted, and long-dashed lines represent 50-, 70-,
120-, and 150-$\mu$m surveys, respectively.}
\end{fv}
%
%
\begin{fv}{2a}
{7cm}
{The cumulative number count of the IRIS galaxies at $B$ band
(per square degree).
The dotted and
dashed lines represent number of normal and starburst populations,
respectively. The total number count is also shown (the solid
line).}
\end{fv}
\begin{fv}{2b}
{7cm}
{The same as figure 2a but at $H$ band.}
\end{fv}
\begin{fv}{3a}
{7cm}
{The cumulative number counts of the starburst population in various
ranges of redshift at $B$ band. The solid line represents the number
count of starbursts in the redshift range of $0<z<0.2$,
the dotted line $0.2<z<1$, the dashed line $1<z<3$, and the
long-dashed line $3<z$.}
\end{fv}
\begin{fv}{3b}
{7cm}
{The same as figure 3a but at $H$ band.}
\end{fv}
\begin{fv}{4a}
{7cm}
{The cumulative number count of the starburst population with our
luminosity-evolution model (solid line). The dashed line shows the
count without evolution.}
\end{fv}
\begin{fv}{4b}
{7cm}
{The same as figure 4a but at $H$ band.}
\end{fv}
\begin{fv}{5a}
{7cm}
{The number counts of the starburst population with evolution in
various ranges of redshift at $B$ band (the meaning of the lines are
the same as figure 3a).}
\end{fv}
\begin{fv}{5b}
{7cm}
{The same as figure 5a but at $H$ band.}
\end{fv}

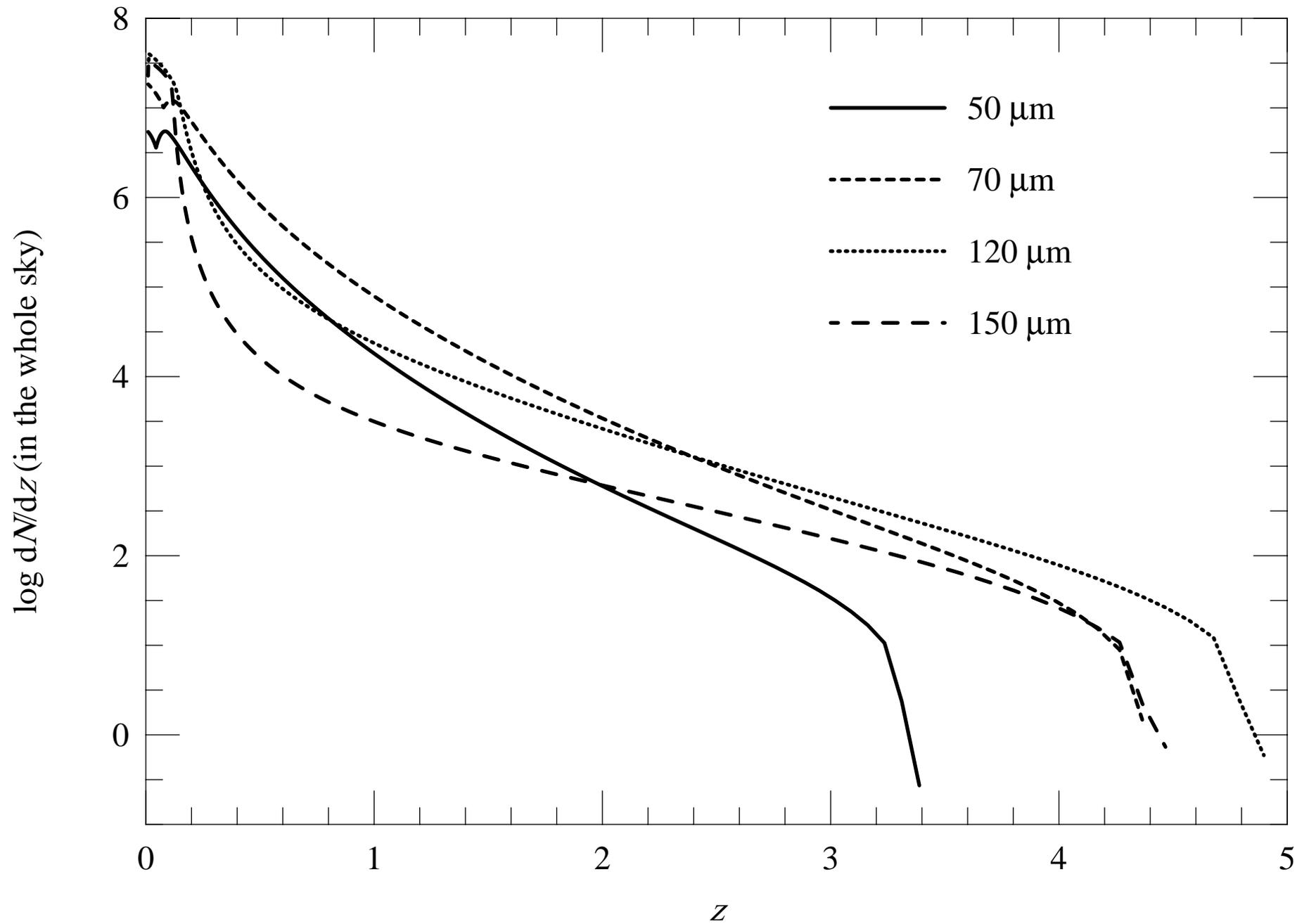
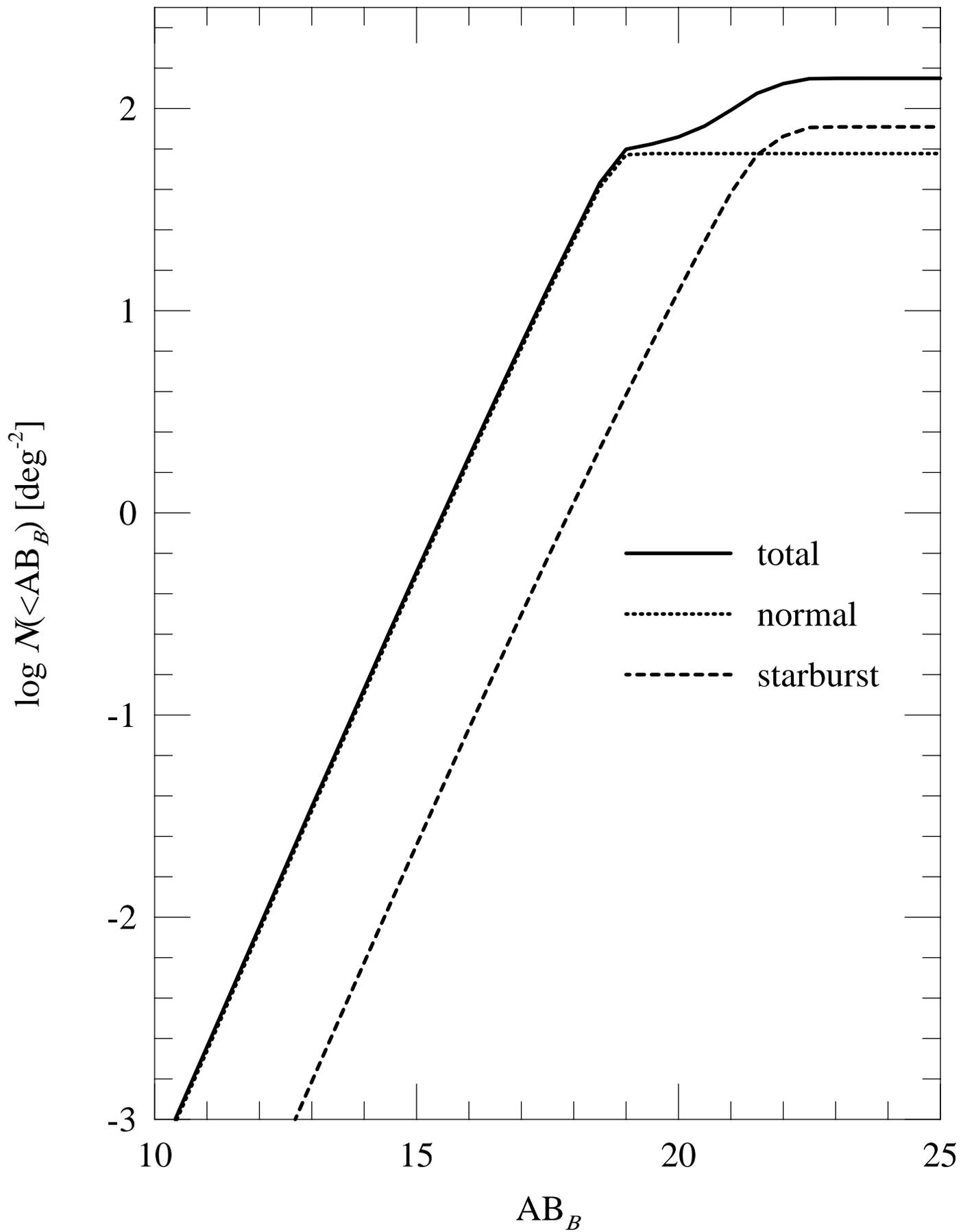
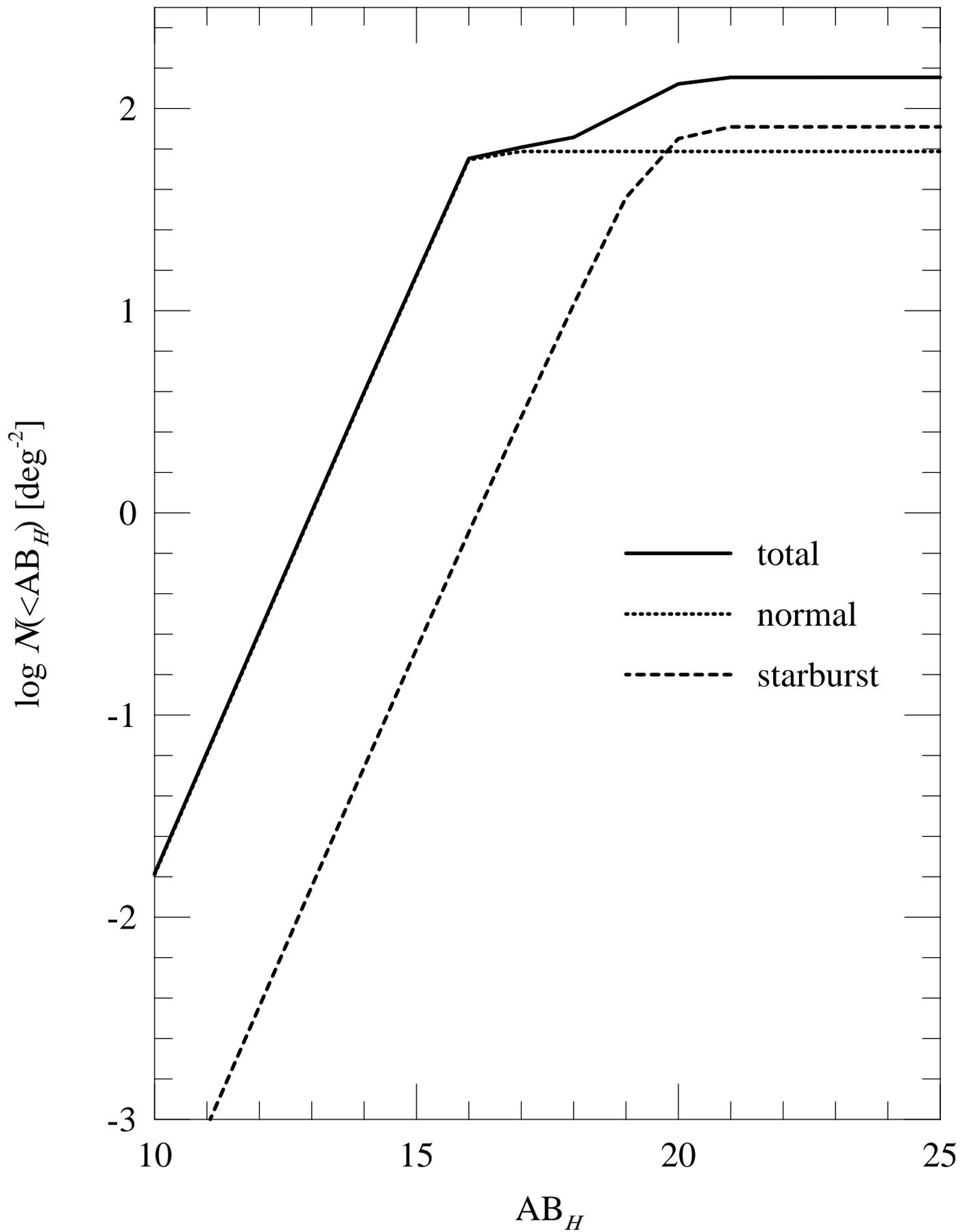
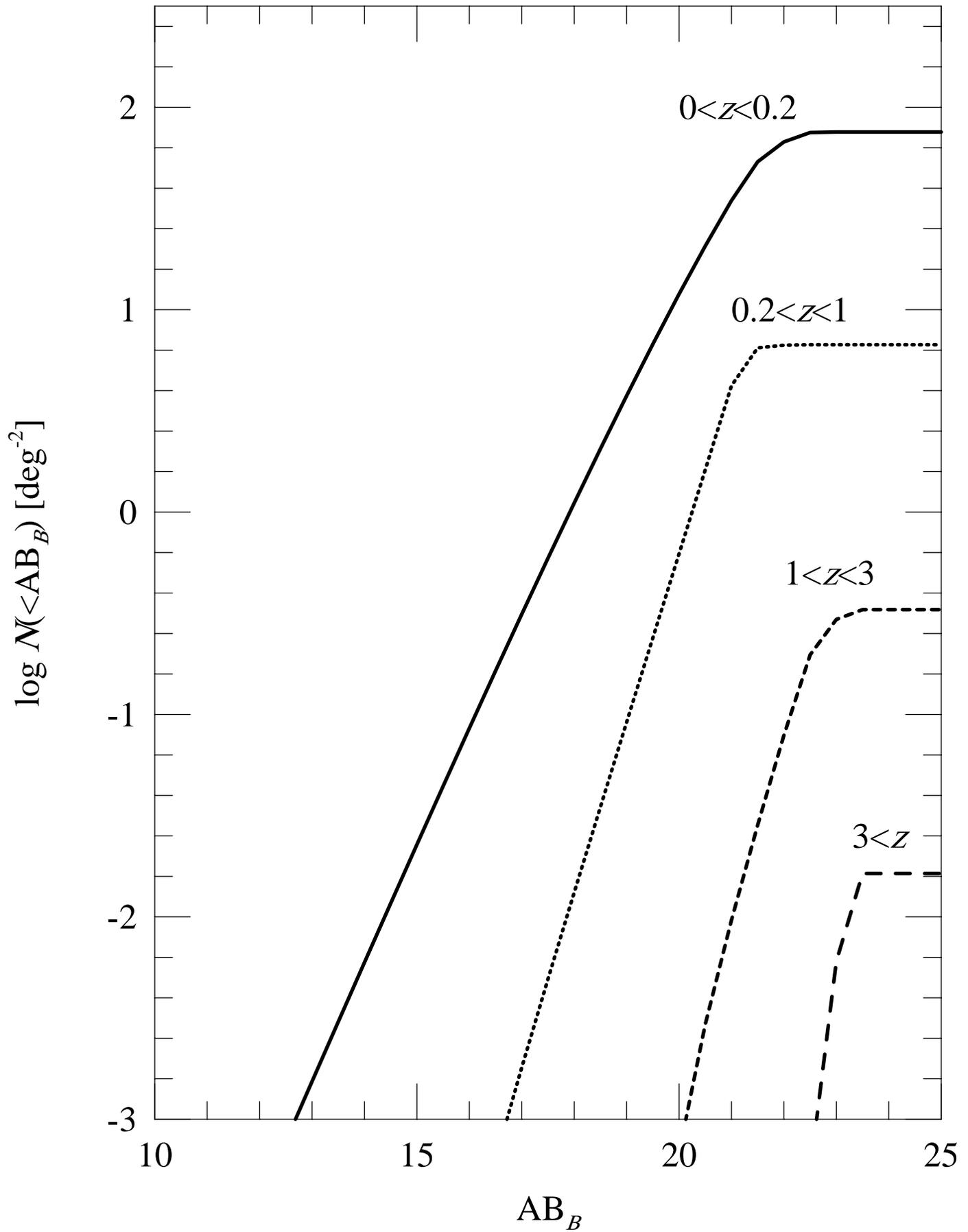
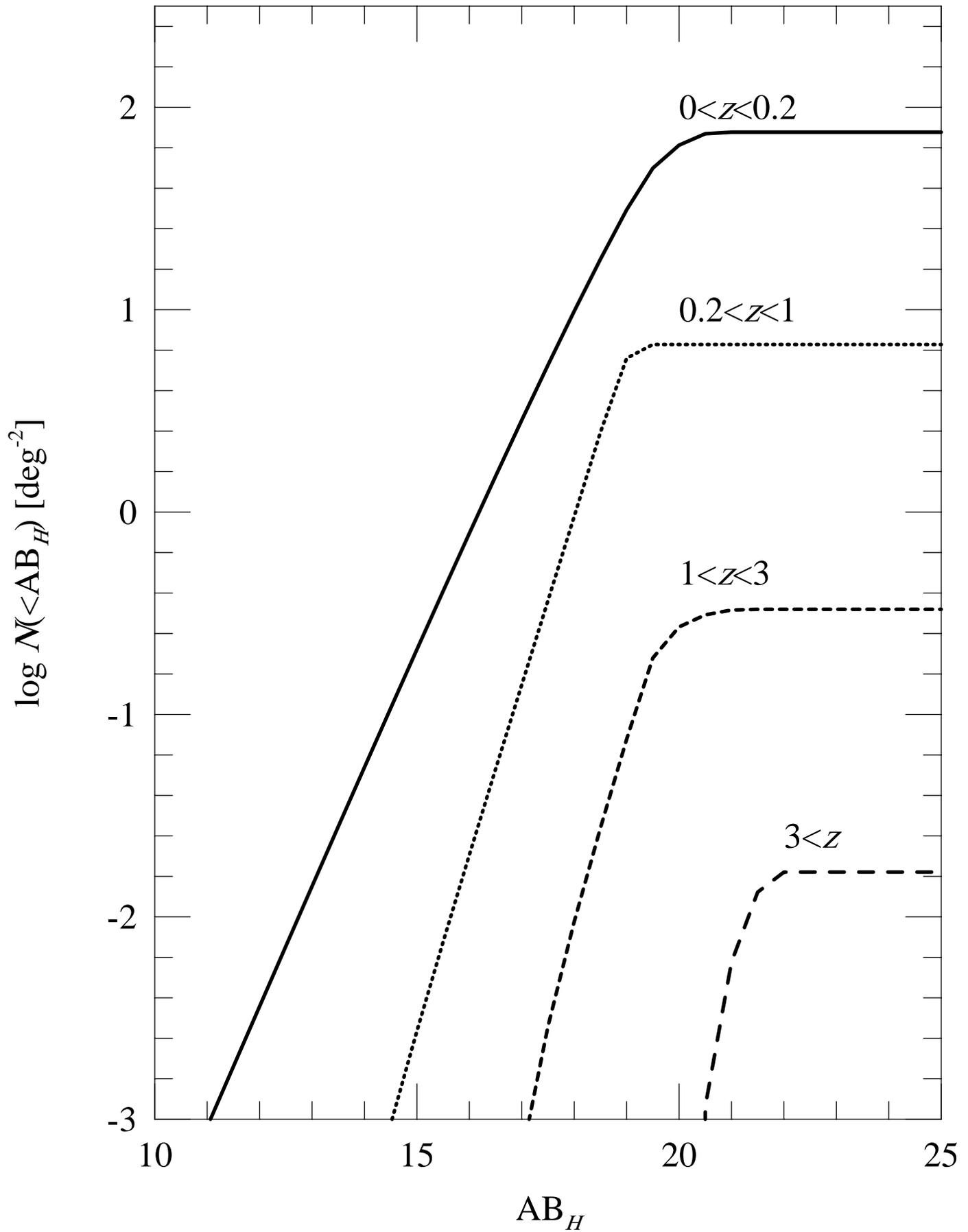
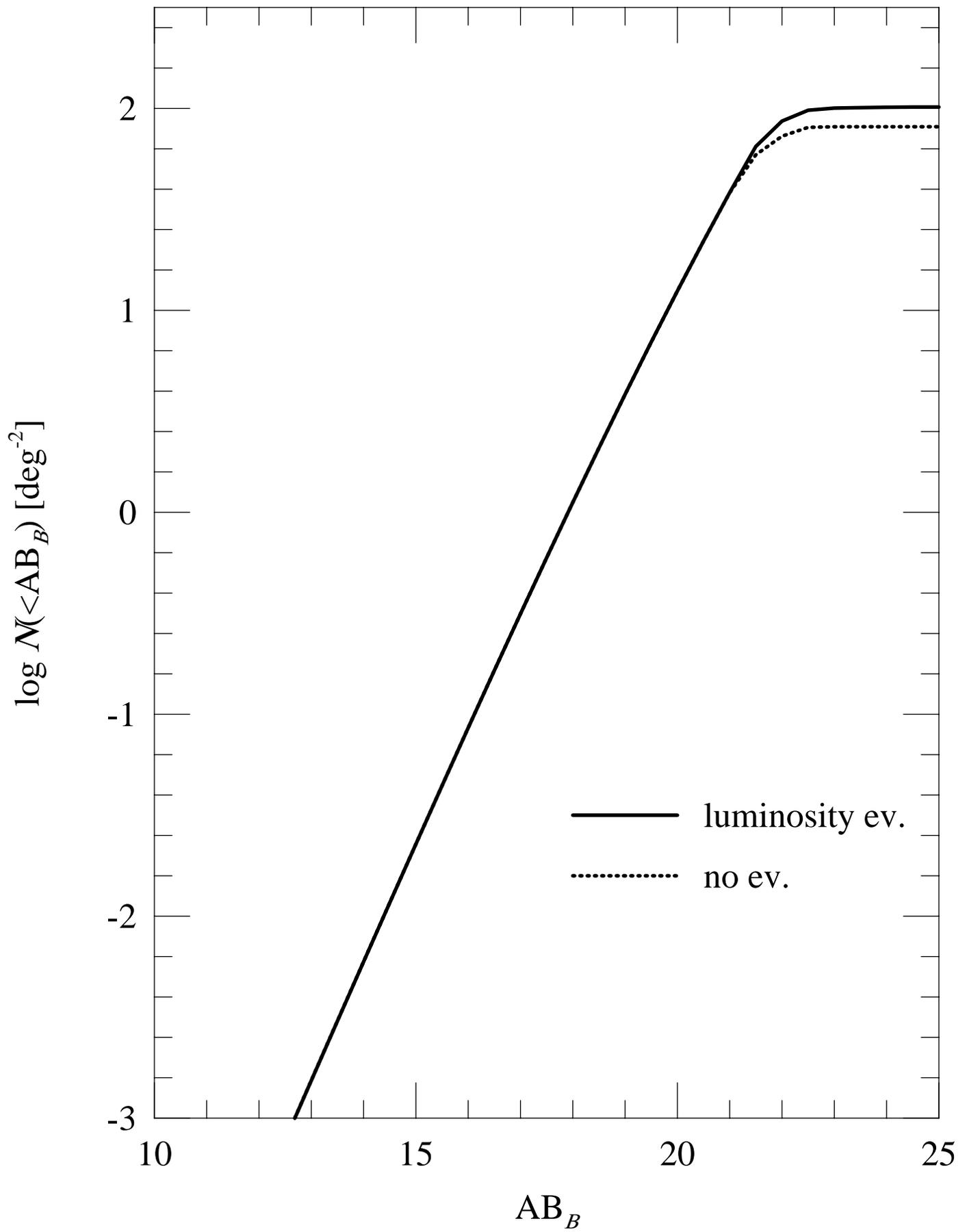
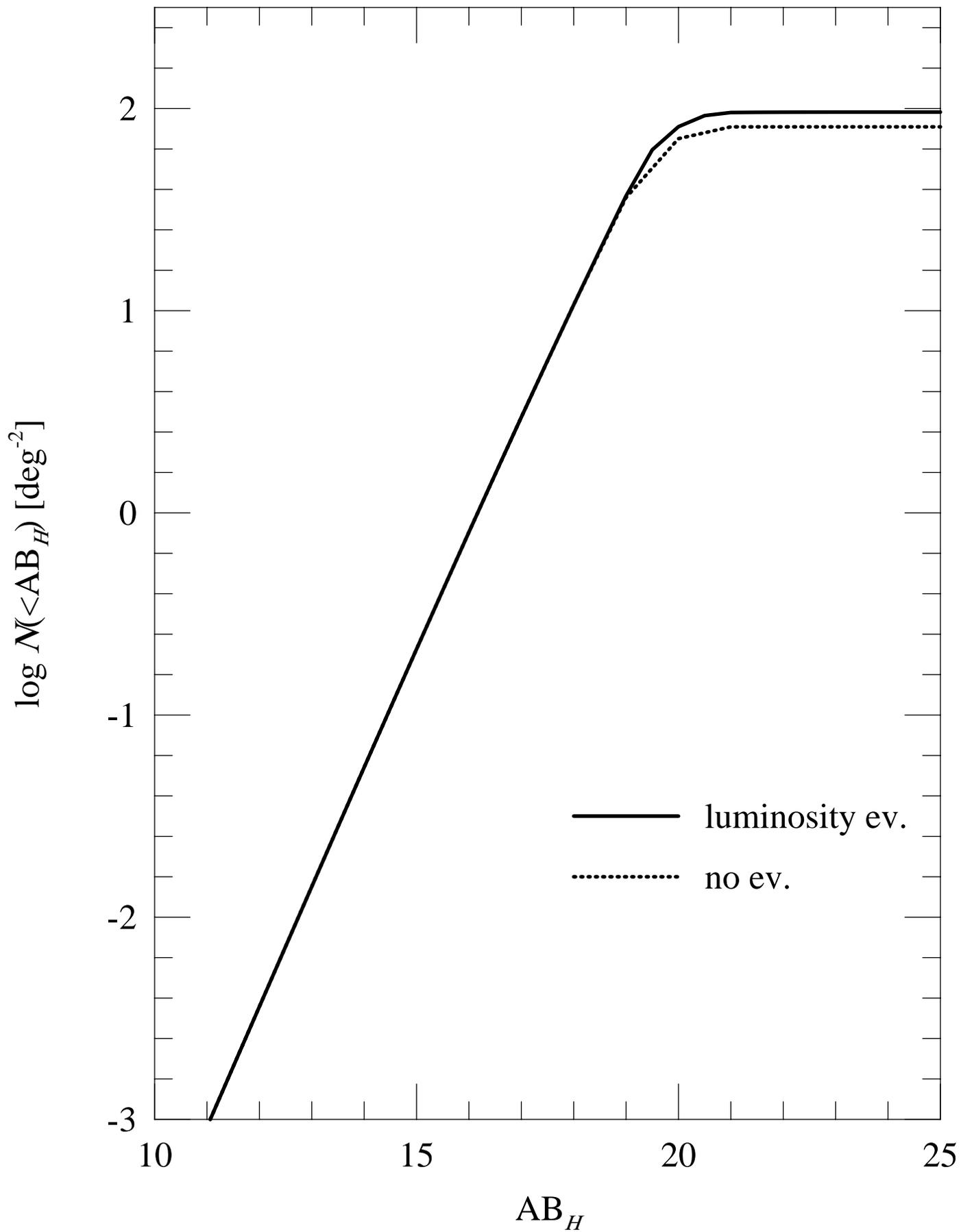
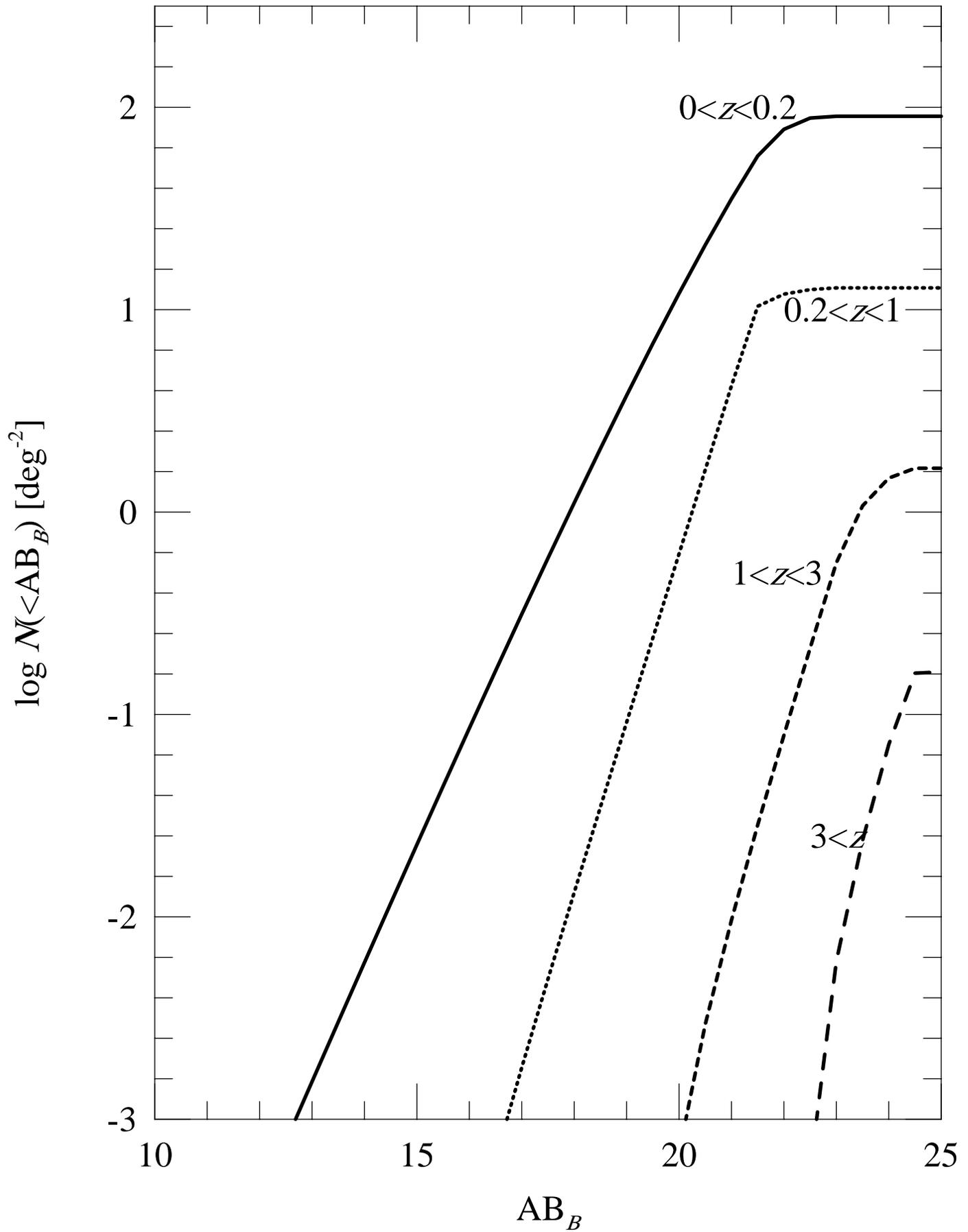
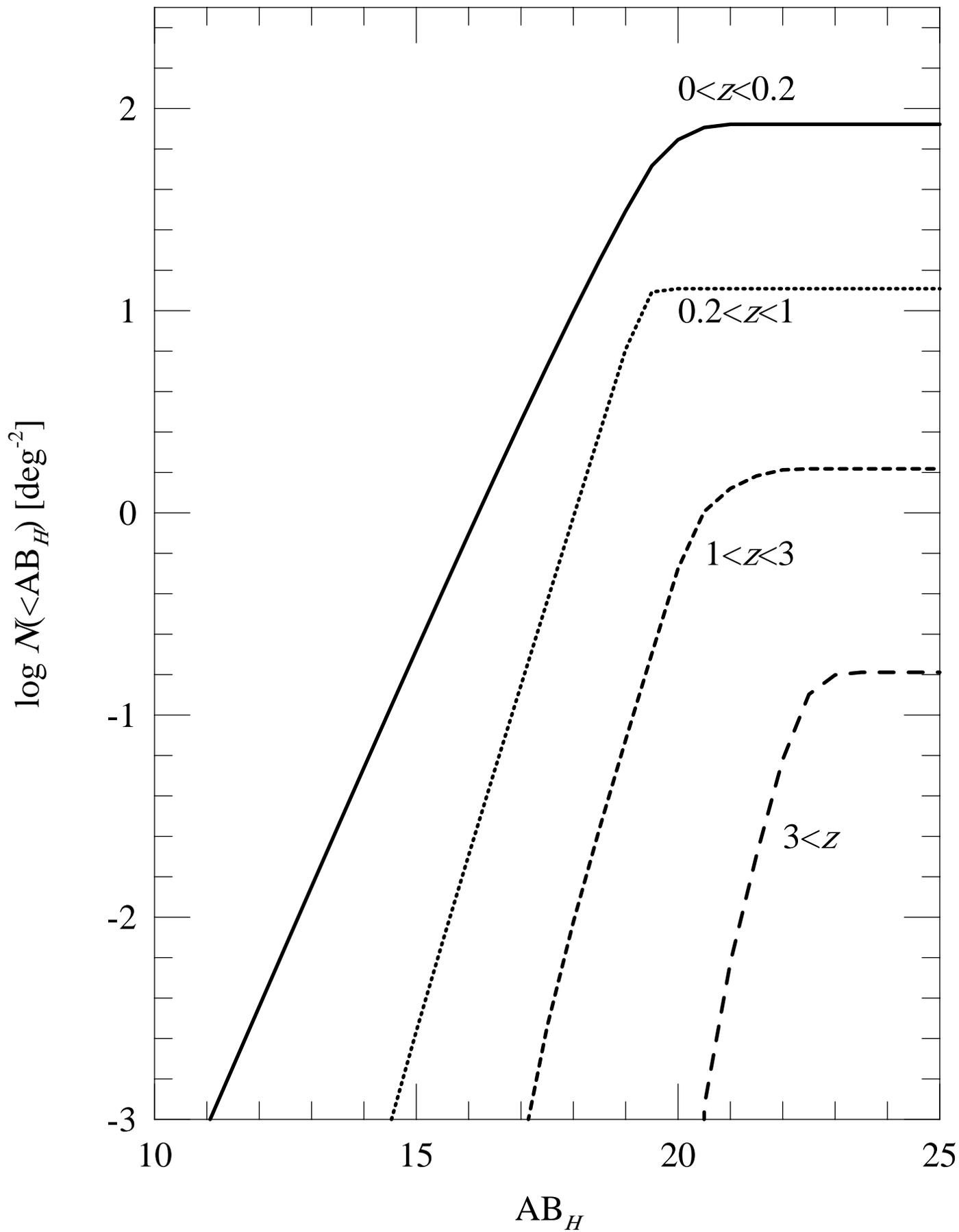
\end{document}